\documentclass[prx,twocolumn,superscriptaddress,floatfix,nopacs]{revtex4}
\usepackage{graphicx,amsfonts,amssymb,amsmath,hyperref,enumerate,soul,pbox,multirow}
\usepackage[normalem]{ulem}
\usepackage{tcolorbox}

\newif\ifhyper
\hypertrue
\ifhyper
\hypersetup{
   citecolor = {green},
   colorlinks = {true}, % false
   urlcolor = {blue} % magenta
}
\fi

\newcommand{\beq}{\begin{equation}}
\newcommand{\eeq}{\end{equation}}
\newcommand{\beqa}{\begin{eqnarray}}
\newcommand{\eeqa}{\end{eqnarray}}
\newcommand{\ket} [1] {\vert #1 \rangle}

\def\ket#1{\vert#1\rangle}

\def\Longarrow{\protect\@lra}
\def\@lra{\relbar\joinrel\relbar\joinrel\relbar\joinrel%
          \relbar\joinrel\rightarrow}

\begin{document}

\title{Tensor networks for complex quantum systems}

\author{Rom\'an Or\'us}
\affiliation{Donostia International Physics Center, Paseo Manuel de Lardizabal 4, E-20018 San Sebasti\'an, Spain}
\affiliation{Ikerbasque Foundation for Science, Maria Diaz de Haro 3, E-48013 Bilbao, Spain}

\begin{abstract}

Tensor network states and methods have erupted in recent years. Originally developed in the context of condensed matter physics and based on renormalization group ideas, tensor networks lived a revival thanks to quantum information theory and the understanding of entanglement in quantum many-body systems. Moreover, it has been not-so-long realized that tensor network states play a key role in other scientific disciplines, such as quantum gravity and artificial intelligence. In this context, here we provide an overview of basic concepts and key developments in the field. In particular, we briefly discuss the most important tensor network structures and algorithms, together with a sketch on advances related to global and gauge symmetries, fermions, topological order, classification of phases, entanglement Hamiltonians, AdS/CFT, artificial intelligence, the 2d Hubbard model, 2d quantum antiferromagnets, conformal field theory, quantum chemistry, disordered systems, and many-body localization.   

\end{abstract}

\maketitle

\section{Introduction}

Tensor network (TN) states \cite{tn} are representations of complex quantum states in terms of { tensor building blocks that can sometimes be regarded as fundamental}. This is somehow similar to { building up a wavefunction with LEGO\textsuperscript{\textregistered} pieces.}  Tensors are the bricks building up the quantum state, and quantum entanglement plays the role of the glue amongst the different pieces. { The individual tensors codify the key properties of the overall wavefunction, intuitively similar to how DNA codifies the key properties of a human being.} This is a naive insight but also very powerful: not only has it helped us to reproduce the theoretical properties of complex quantum systems, but also to develop new numerical simulation algorithms that go well beyond the limits of what could be simulated a few years ago. It helps us in so many ways that the field of tensor networks  has become very active and interdisciplinary. Our purpose with this review is to provide an entry-point text on key aspects and recent developments, pinpointing important references.  

\subsection{Some history} 

One does not wake up one day and think all of a sudden about the TN structure of complex quantum states. Instead, TNs have evolved progressively over the years. Kramers and Wannier proposed an approximation to the 2-dimensional (2d)  classical Ising model \cite{KW} that is considered as a precursor of TN variational methods. Later on, Baxter \cite{baxter} made use of similar ideas when dealing with classical partition functions. But it was around the 90's when people investigated more seriously how such structures were intimately linked to wave-functions of quantum lattice systems. In particular, a key development was that of Density Matrix Renormalization Group (DMRG) \cite{dmrg} by White. DMRG was constructed as a technique to keep the relevant degrees of freedom in a renormalization procedure targeting low-energy eigenstates of 1d Hamiltonians. White understood that the relevant degrees of freedom to be kept are, in fact, \emph{entanglement} degrees of freedom of the wave-function. Later on, it was proven that the state produced by DMRG is a ``finitely-correlated state", i.e., essentially what we call today Matrix Product State (MPS) \cite{MPS}, or equivalently, what some mathematicians call Tensor Train \cite{TT}. { Moreover, it was also understood that DMRG was a variational optimization algorithm over MPS \cite{ostlund}.} The success of DMRG was explosive, and the method quickly became the tool of reference for 1d quantum lattice systems at low energies. 

A second wave of results came around the 2000's. Physicists working on quantum information theory started to wonder about entanglement in low-energy eigenstates of Hamiltonians, and its behaviour in quantum phase transitions. Developments such as the entanglement entropy of a block for 1d quantum critical systems \cite{cft} helped to understand that entanglement had a \emph{structure}. Moreover, it was also discovered that MPS are particularly well-suited for gapped 1d quantum lattice systems with local interactions \cite{mpsgapped}, precisely due to their entanglement structure. Subsequent developments generalized this to other scenarios: higher dimensional systems with Projected Entangled Pair States (PEPS) \cite{PEPS}, critical systems with the Multiscale Entanglement Renormalization Ansatz (MERA) \cite{mera}, and more. 

From then on, developments on TN methods continued mostly at the crossover of condensed matter physics and quantum information. But in addition to this, it came as a pleasant surprise that TNs were also relevant in other scientific areas. Examples are quantum gravity, where the MERA { has been proposed to be linked to geometry of space, e.g.. via the Anti-de Sitter / Conformal Field Theory correspondence (AdS/CFT)} \cite{swingle}, and artificial intelligence, where it has been proven that neural networks have a TN structure \cite{NNetTN}. And more applications are being found every day, sometimes in the most unexpected places. Essentially, anywhere that there is a structure of correlations, there is also a TN, meaning that there is room to apply our knowledge on quantum many-body entanglement. 

\subsection{Some maths} 

\begin{figure}
\centerline{\includegraphics[width=0.7\columnwidth]{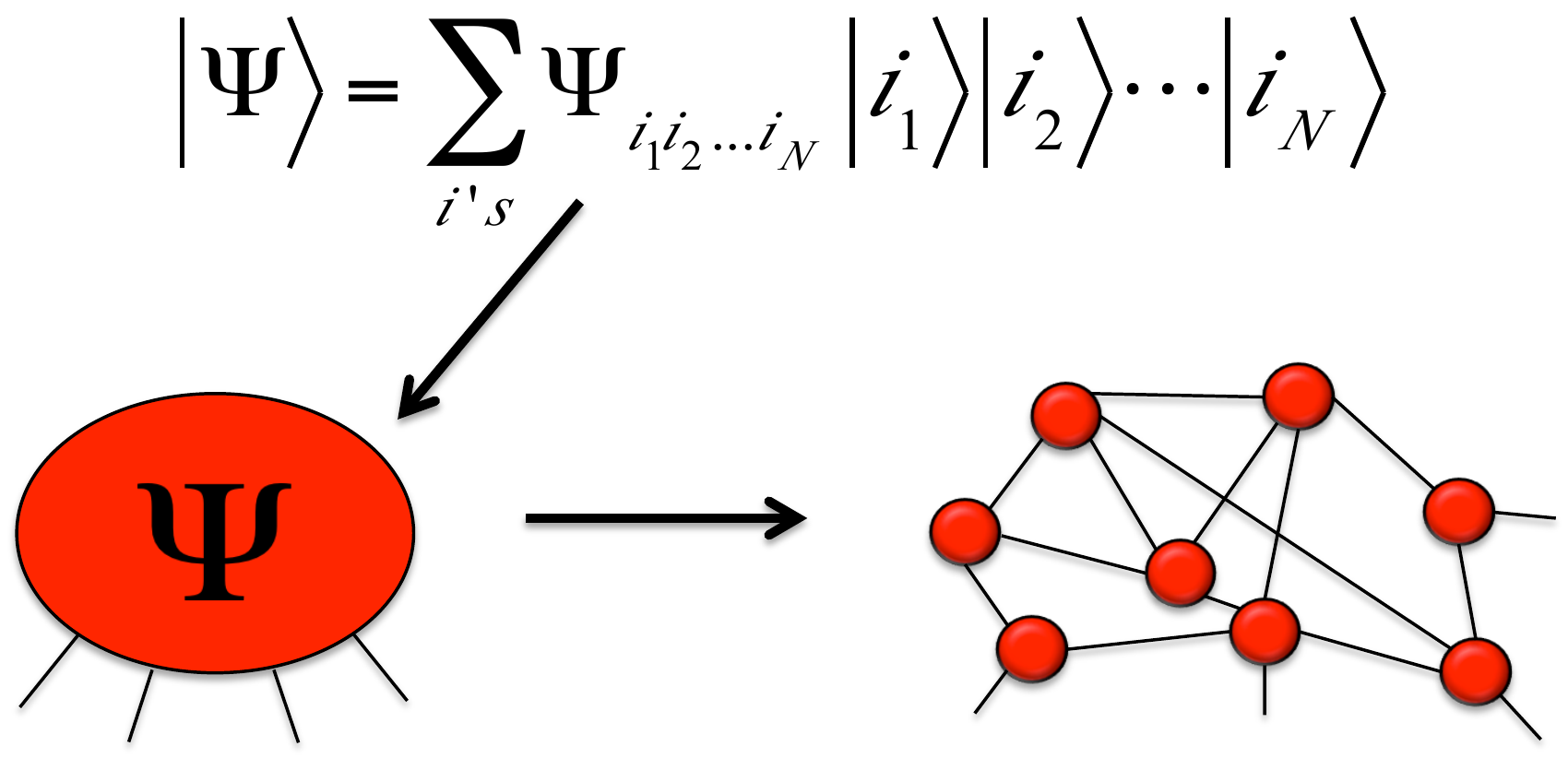}}
\caption{The coefficient of the quantum state of a many-body system with $N$ sites (say, $N$ spins) can be understood as a tensor with exponentially many coefficients in the system's size. The structure of this tensor is accounted for by a tensor network, which is a network of tensors interconnected by ancillary indices that take into account the structure and amount of entanglement in the quantum many-body state. This is represented here diagrammatically, where shapes correspond to tensors, lines to indices, and lines connecting shapes to contracted (summed) common indices. { The tensor network on the right hand side is intended to be ``generic". More useful examples of tensor networks are discussed in Fig.~\ref{fig1}.}}
\label{fig0}
\end{figure}

TNs are representations of quantum many-body states based on their local entanglement structure. { They arise naturally whenever one has a tensor product structure (more generally, a tensor category or a braided fusion category).} Take for instance a quantum many-body system of $N$ spins-$1/2$. Any wave-function of the system can be described, in a computationally inefficient way, in terms of $O(2^N)$ coefficients once an individual basis for spins is chosen. As such, these coefficients can be understood as a tensor with $N$ indices, where each index takes two possible values (say, spin ``up" and ``down"). { For our purposes, a tensor will be, simply, a multidimensional array of complex numbers.} A plausible motivation for TNs is that one can replace this tensor by a network of interconnected tensors with less coefficients, { see Fig.~\ref{fig0} for a generic case and Fig.~\ref{fig1} for useful examples.} This construction defines a TN, and it depends on $O({\rm poly}(N))$ parameters only, assuming that the rank of the interconnecting indices is upper-bounded { by a constant $\chi$ (sometimes also called $D$, as we shall see later} \footnote{{ Depending on the situation one may use $\chi$, or $D$, or \emph{both}. For instance, we use $\chi$ for the bond dimension of MPS and MERA. However, when dealing with 2d tensor networks such as a PEPS, one uses $D$ for the PEPS bond dimension, and reserves $\chi$ for the bond dimension of the environment tensors of 2d numerical algorithms, see Sec.~\ref{thissec}.}}). This constant is called  ``bond dimension". Similarly, interconnecting indices in the network are also called ``bond indices". In practice they provide the structure of the many-body entanglement in the quantum state, and parameter $\chi$ turns out to be a quantitative measure of the entanglement present in the quantum state. As an example, $\chi=1$ corresponds to a separable product state (i.e., mean field theory), whereas any $\chi > 1$ provides non-trivial entanglement properties. In addition, it is possible to see that  TN states satisfy the so-called Òarea-lawÓ for the entanglement entropy \cite{arealaw}, and characterize the relevant corner of the Hilbert space for a quantum many-body systems at low energies.

A key mathematical ingredient in numerical algorithms dealing with TNs is the Singular Value Decomposition (SVD), which is strongly tied to the Schmidt decomposition of quantum states. This is specified in the Technical Box. The Schmidt decomposition states that it is possible to write a bipartite quantum state in terms of orthonormal correlated basis for the two parties (the Schmidt basis), together with $\chi$ real and positive coefficients (the Schmidt coefficients). Parameter $\chi$ is called the Schmidt rank. At the level of the coefficients of the wave-function, { if \emph{orthonormal}  basis for the $A / B$ vector spaces are considered (see the box), then} the Schmidt decomposition is equivalent to the Singular Value Decomposition of the matrix containing the coefficients, and Schmidt coefficients correspond to the singular values. As we shall see, the SVD is one of the key tools in TN algorithms. The interested reader is referred to, e.g., Ref.~\cite{tn} for more details. 

{ Another important concept is that of \emph{canonical form}. Mathematically, the canonical form of a TN is that in which all the bond indices simultaneously correspond to orthonormal Hilbert spaces. For a TN without loops this can be achieved by playing with sequential SVDs, see, e.g., Refs.~\cite{tebd, itebd, canonical, tn} for more details. For TNs with loops, however, this is formally not possible. On top of providing a useful representation of the TN, the canonical form helps a lot in simplifying tensor contractions as well as in providing truncation schemes in TN algorithms \cite{tebd}.} 

\section{Main tensor network structures}

\begin{figure}
\centerline{\includegraphics[width=0.85\columnwidth]{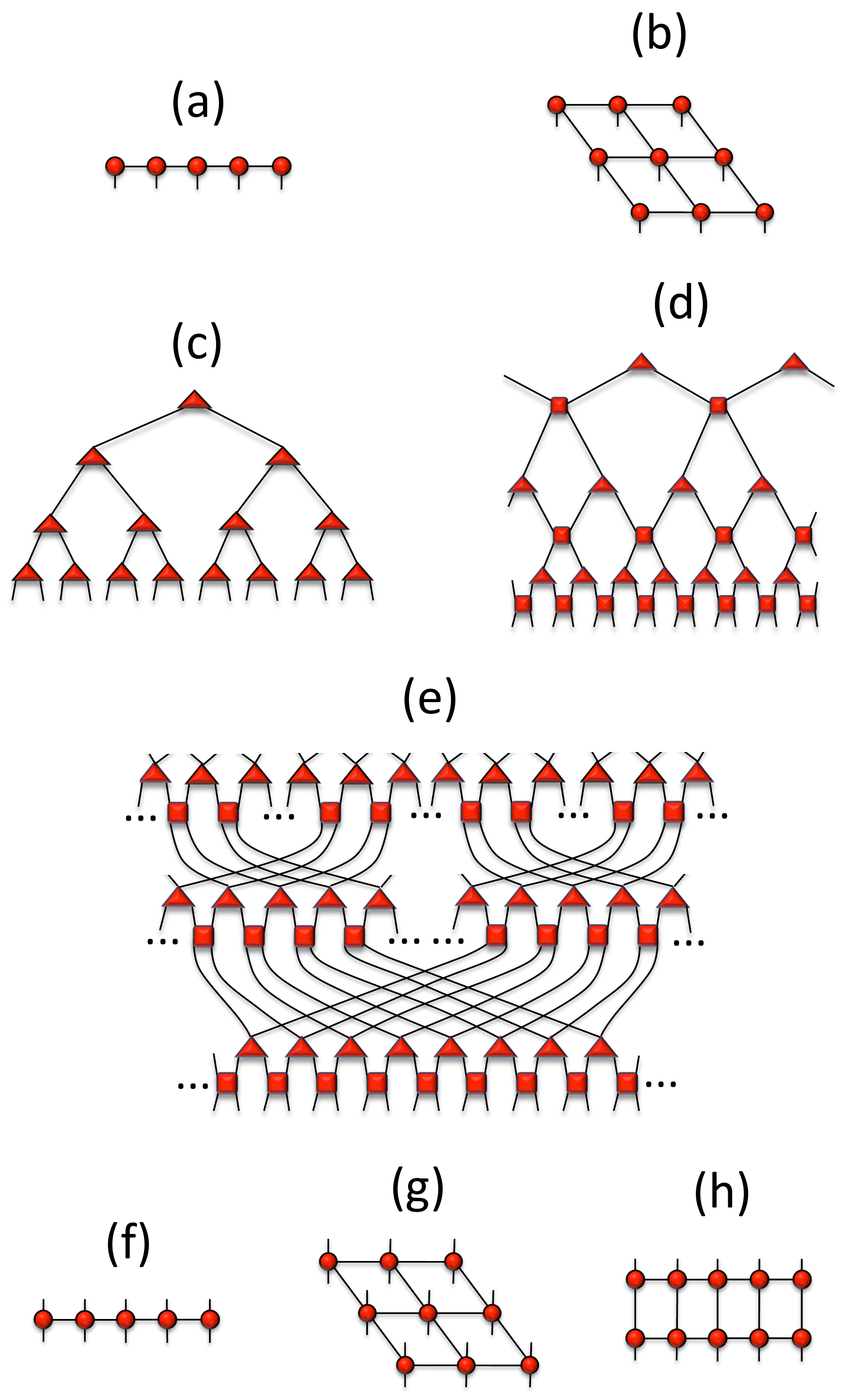}}
\caption{Several tensor networks in diagrammatic notation. (a) Matrix Product State (MPS). (b) Projected Entangled Pair State (PEPS) for a  square lattice. (c) Tree Tensor Network (TTN). (d) Multiscale Entanglement Renormalization Ansatz (MERA). (e) Branching MERA. (f) Matrix Product Operator (MPO). (g) Projected Entangled Pair Operator (PEPO). (h) Matrix Product Density Operator (MPDO), where tensors in the upper row are the hermitian conjugates (with respect to up/down indices) of the tensors in the lower row, so that the whole operator is hermitian and positive.}
\label{fig1}
\end{figure}

Let us now introduce some of the most important TN structures. These are shown diagrammatically in Fig.~\ref{fig1}, and a comparison of their properties is provided in Table \ref{tab1}. In what follows we sketch briefly their motivation and main properties. 

\bigskip
\emph{(i) Matrix Product States (MPS) \cite{MPS}:} these are 1d arrays of tensors, as in Fig.~\ref{fig1}(a). Generally speaking, they correspond to low-energy eigenstates of gapped 1d local Hamiltonians \cite{mpsgapped}. They also satisfy a 1d area law for the entanglement entropy of a block (see Table \ref{tab1}). Expectation values of local observables can be efficiently computed, and have a finite correlation length. Hence, MPS cannot formally represent the entanglement structure of a quantum critical system. However, they are extremely efficient to manipulate, { and over the years people have used a rich variety of techniques based on finite-size scaling and finite-entanglement scaling \cite{fes} to use them in order to extract properties of 1d quantum critical systems, see Ref.~\cite{zou} for a recent example.}

\bigskip 
\emph{(ii) Projected Entangled Pair States (PEPS) \cite{PEPS}:} these are 2d arrays of tensors, as in Fig.~\ref{fig1}(b) for the particular case of a square lattice. They are known to capture the correct correlation structure of low-energy eigenstates of 2d local Hamiltonians satisfying the 2d area law as well as of 2d thermal states \cite{mpsgapped, PEPSground}. Unlike MPS, PEPS can handle critical correlation functions \cite{PEPScrit}, but cannot be contracted both efficiently and exactly \cite{PEPSNPHard}. This is the reason why people developed approximate methods to manipulate them, some of which are sketched later in this review. They can also handle topological order, both chiral \cite{chiralPEPS} and non-chiral \cite{StringnetPEPS}. Still, it is not fully understood if they can handle chiral topological order with gapped bulk excitations. 

\bigskip 
\emph{(iii) Tree Tensor Networks (TTN) \cite{TTN}:} these are tree-like structures as in Fig.~\ref{fig1}(c). By construction, TTNs have a finite correlation length and an entropy that, on average, satisfies a 1d area law. They are therefore well suited for gapped 1d systems, though they have also been used quite extensively to deal with 1d critical systems \cite{1dTTN}, as well as 2d systems \cite{2dTTN}. By construction, they codify a Wilsonian renormalization structure, if the tensors codify a coarse-graining of the Hilbert space and are therefore isometries.  

\bigskip
\emph{(iv) Multiscale Entanglement Renormalization Ansatz (MERA) \cite{mera}:} these are structures such as those in  Fig.~\ref{fig1}(d), where we show the case of a 1d MERA. They are essentially like a TTN of isometries, but including the so-called ``disentanglers", which are unitary operators that account for entanglement amongst neighbouring sites. Thus, MERAs are made from unitaries and isometries, and have a number of remarkable properties. For instance, they can handle the entanglement entropy of critical 1d systems \cite{MERAcrit}. Moreover, they are efficiently contractable. They have an extra holographic dimension that encodes a renormalization scale, related to the so-called Entanglement Renormalization \cite{ER}. Finally, MERA is believed to be linked to the AdS/CFT correspondence in quantum gravity \cite{swingle}.  

\bigskip 
\emph{(v) Branching MERA (bMERA) \cite{branmera}:} this is shown in Fig.~\ref{fig1}(e). It is similar to the MERA, with the additional fact that at every renormalization scale, the MERA decouples into several copies. In this way one can have an \emph{arbitrary} scaling of the entanglement entropy of a block, thus allowing to reproduce the entanglement structure of systems that violate the area-law. Physically speaking, this structure accounts for the decoupling of degrees of freedom at different renormalization scales (e.g.,  spin-charge separation in electronic solid-state systems). 

\bigskip 
\emph{(vi) Operators and mixed states:} operators can be conveniently represented by TNs, e.g., by Matrix Product Operators (MPO) in 1d \cite{MPO} and Projected Entangled Pair Operators (PEPO) in 2d \cite{PEPO}, see Fig.~\ref{fig1}(f,g). Mixed states can also be represented by such structures, and also by purification-like structures such as Matrix Product Density Operators (MPDO) in 1d \cite{MPDO}, see Fig.~\ref{fig1}(h). The advantage of MPDOs is that they are positive by construction. However, the amount of correlations that they can carry for a fixed bond dimension is typically lower than that of a generic MPO with similar bond dimension \cite{Gemma}.
\begin{tcolorbox}[standard jigsaw, opacityback=0]
{\bf Schmidt \& Singular Value Decompositions}

\hrulefill

\bigskip 

The Schmidt Decomposition (SD) states that a bipartite quantum state $\ket{\Psi}_{AB} \in \mathcal{H}_A \otimes \mathcal{H}_B$ can always be written as 
\beq
\ket{\Psi}_{AB} = \sum_{\alpha = 1}^\chi \lambda_\alpha \ket{\alpha}_A \ket{\alpha}_B, 
\label{sd}
\eeq
with $\{Ê\ket{\alpha}_A \}, \{Ê\ket{\alpha}_B \}$ orthonormal basis (the \emph{Schmidt basis}) respectively for Hilbert spaces $\mathcal{H}_A$ and $\mathcal{H}_B$, $\lambda_\alpha > 0$ (the \emph{Schmidt coefficients}), and  $\chi \le {\rm min}(d_A, d_B)$ (the \emph{Schmidt rank}) with $d_{A/B}$ the dimension of $\mathcal{H}_{A/B}$.  Proving the SD is easy using the Singular Value Decomposition (SVD): write the state as $\ket{\Psi}_{AB} = \sum_{ij} \Psi_{ij} \ket{i}_A \ket{j}_B$ in some arbitrary basis for $A$ and $B$. The SVD of the matrix $\Psi$ of coefficients reads $\Psi = U \Lambda V^\dagger \nonumber$, with $U^\dagger U = {\mathbb I}_{\chi}$, $V^\dagger V = {\mathbb I}_{\chi}$ (i.e., they are \emph{isometries}), and $\Lambda$ a $\chi \times \chi$ diagonal matrix with real positive entries (the \emph{singular values}). In terms of matrix components, this reads
\beq
\Psi_{ij} = \sum_{\alpha, \beta = 1}^{\chi} U_{i \alpha} \Lambda_{\alpha \beta} \left( V^\dagger \right)_{\beta j} 
= \sum_{\alpha = 1}^\chi  U_{i \alpha} \lambda_{\alpha} \left( V^\dagger \right)_{\alpha j}, \nonumber  
\eeq
or in diagrammatic form
\newcommand{\diagram}[2]{\,\vcenter{\hbox{\includegraphics[scale=0.25,page=#2]{./#1.pdf}}}\,}
\begin{equation}
\diagram{figsvd}{1}, \nonumber
\end{equation} 
where we used $\Lambda_{\alpha \beta}Ê= \lambda_\alpha \delta_{\alpha \beta}$ and the fact that there are at most $\chi$  non-zero singular values $\lambda_\alpha$. Inserting the above equation into $\ket{\Psi}_{AB}$ and rearranging the sums produces  the SD in Eq.(\ref{sd}). 

The conclusion from this small exercise is that the SD for a quantum state is the same as the SVD for its coefficients, { provided that one has orthonormal basis for the Hilbert spaces $\mathcal{H}_{A/B}$.} This has important consequences in approximation  schemes of TN methods: truncating in singular values is related to truncating in Schmidt coefficients, { either exactly or approximately (as in the case of non-orthonormal basis for $\mathcal{H}_{A/B}$).} And this is linked to truncating in the entanglement of the quantum state, since the entanglement entropy $S$ between $A$ and $B$ satisfies   
\beq
S = - \sum_{\alpha = 1}^\chi \lambda_\alpha^2 \log \lambda_\alpha^2 \le \log \chi .\nonumber 
\eeq

\label{box1}
\end{tcolorbox}

\bigskip 
\emph{(vii) Continuous TNs:} finally, we would like to mention that the structures above allow for a continuum limit, therefore becoming an ansatz for low-energy functionals of quantum field theories, as well as to operators on those functionals. For instance, one has continuous MPS (cMPS) \cite{cMPS}, continuous MERA (cMERA) \cite{cMERA} and continuous PEPS (cPEPS) \cite{cPEPS}. While cMPS and cMERA have been used for a variety of applications already, cPEPS are still to be very much explored.

\begin{table}
	\centering
	\begin{tabular}{||c||c|c|c|c|c||} 
	\hline 
  & MPS & 2d PEPS & TTN & 1d MERA & 1d bMERA  \\
         \hline 
         \hline
  $S(L)$ & $O(1)$ & $O(L)$ & $O(1)$ & $O( \log L)$ & $O(L)$ \\
  $ \langle O \rangle$  & exact & approx. & exact & exact & exact \\
  $\xi$ & $ < \infty$  &  $ \le \infty$ & $ < \infty$ & $ \le \infty$ & $ \le \infty$ \\
  Tensors & any & any & any & unit./isom. & unit./isom. \\
  Can. form & obc, $\infty$ & no & yes & -- & -- \\   
  \hline 
         \end{tabular}
         \caption{Comparison of several properties for some of the TNs discussed in the main text: entanglement entropy $S(L)$ of a block of length $L$ (i.e., with $L$ sites in 1d and $L \times L$ in 2d), calculation of a local expectation value $\langle O \rangle$, correlation length $\xi$, constraints on tensors, and exact canonical form. MERA and bMERA are built from unitary and isometric tensors, and we therefore do not consider a canonical form for them. For MPS, ``obc" stands for ``open boundary conditions".}
                  \label{tab1}
\end{table}

\section{Main algorithms} 

Let us now sketch very briefly the main ideas behind some of the most important numerical algorithms using TNs. Our purpose here is to explain the basic idea behind key families of numerical TN methods, leaving out technical details about implementation. Readers interested in more information are addressed to the specific papers explaining the details of each technique. Let us also stress that numerical TN methods are also difficult to classify according to a unique criteria. Here we do it as follows: we first introduce methods to obtain 1d states, then methods to contract 2d TNs, then methods to obtain 2d states, and finally mixed methods where TNs are combined with other techniques.  

\subsection{Methods to obtain 1d states} 

\bigskip 
\emph{(i) Density Matrix Renormalization Group (DMRG) \cite{dmrg}:} by far the most famous TN method. DMRG was originally proposed as a renormalization procedure over the ground-state wave-function of a 1d quantum lattice system. In the modern formulation, it is a variational optimization algorithm over the family of MPS \cite{uli}. In particular, one sweeps back and forth over the different tensors of an MPS, with tensor coefficients treated as variational parameters in order to minimize the expectation value of a given Hamiltonian. At every step, the optimization can be written as a quadratic problem for the tensor parameters, which can be solved via standard linear algebra. The method also makes use of other tricks, such as the canonical form for open boundary conditions in order to improve stability and performance \cite{uli}. It can also be adapted to translationally-invariant systems in the thermodynamic limit, the so-called infinite-DMRG (iDMRG) \cite{idmrg}. There are also extensions of the method to deal with periodic boundary conditions \cite{pbc} as well as low-energy excitations \cite{exmps, natalia}. For open boundary conditions and infinite-size systems the computational cost is $O(\chi^3)$, with $\chi$ the MPS bond dimension. For periodic boundary conditions, if no further approximations are introduced the computational cost is $O(\chi^5)$. 

\bigskip 
\emph{(ii) Time-Evolving Block Decimation (TEBD) \cite{tebd}:} this method, introduced by Vidal in 2004, allows to compute the time evolution of an MPS. If the evolution is in real-time, then it allows to compute dynamics as long as the state remains slightly entangled. If it is in imaginary-time, then it allows to approximate ground states and is a plausible alternative to DMRG. TEBD has also been applied successfully to infinite-size systems (infinite-TEBD) \cite{itebd}. The asymptotic scaling of the computational cost is similar to the one of DMRG, i.e., $O(\chi^3)$ for an MPS of bond dimension $\chi$ with open boundary conditions. TEBD is based on the canonical form of an MPS \cite{canonical} and the truncation of the MPS bond dimension via SVD, which provides an optimal local approximation between MPS formally justified only for TNs without loops. { This truncation scheme is similar to the one in the so-called ``2-site DMRG", where tensors for two sites (and not just one) are optimized by a variational update following an SVD.} 

\bigskip 
\emph{(iii) Tree Tensor Networks (TTN):} in addition to MPS, TTNs are also a useful tool to study 1d systems. Concerning this, the procedures used in DMRG and TEBD have been extended also to TTNs to study 1d gapped and critical systems \cite{TTN, 1dTTN}. This is possible given the absence of loops in the TTN. The computational cost depends on the specifics of the tree, but typically it can be boiled down to $O(\chi^4)$, with $\chi$ the bond dimension of the TTN \cite{TTN}. 

\bigskip 
\emph{(iv) 1d MERA:} the MERA \cite{mera} can also be used as a variational ansatz to approximate ground-state properties. For the case of 1d systems, this has been done both for gapped \cite{mera} and critical \cite{MERAcrit} systems. The variational optimization is however tricker than for, e.g., DMRG, because of the constraints on the tensors (they must be unitaries and isometries). Such optimization can be done by using a number of techniques, as explained in detail in Ref.~\cite{MERAmeth}. Overall, the computational cost depends on the actual structure of the MERA. For instance, for the so-called binary MERA the cost is $O(\chi^9)$, whereas for the so-called ternary MERA the cost is $O(\chi^8)$, and for the so-called modified-binary MERA it is $O(\chi^7)$ \cite{MERAcrit}. 

\bigskip 
\emph{(v) Tangent Space Methods \cite{tangent}:} a whole new family of methods is based on the idea that MPS can be understood as a \emph{manifold}, in the sense of differential geometry. With this in mind, it is possible to obtain a variety of highly-efficient and accurate methods using the concept of \emph{tangent space} of the MPS manifold. A remarkable example is the Time-Dependent Variational Principle (TDVP) algorithm \cite{1dtdvp} for MPS. This algorithm uses concepts of differential geometry to compute the time evolution of an MPS either in real or imaginary-time, without the need of a Trotter decomposition (as in TEBD) and preserving naturally all the spatial symmetries of the physical system. The idea also allows for the computation of low-energy excitations and energy bands by means of an ansatz with well-defined momentum \cite{extdvpmps}. The formalism of the tangent space can also be applied to the variational optimization of the expectation value of a Hamiltonian, the so-called Variational Uniform MPS (VUMPS) algorithm \cite{vumps}. The asymptotic computational cost of this family of methods for 1d systems is similar to the one of other MPS approaches, i.e., $O(\chi^3)$ for an infinite-size MPS of bond dimension $\chi$. 

\bigskip 
\emph{(vi) Other:} people have used other strategies to develop new algorithms focusing on the properties of MPS. Examples are alternative methods to simulate time evolution in 1d \cite{folding}, and to simulate open quantum dynamics, { including non-Markovian processes \cite{oqd0} and effective small reservoirs \cite{oqd}.} 

\subsection{Methods to contract 2d tensor networks} 
\label{thissec}

Here we consider methods to contract a TN without open indices, focusing mostly on methods to compute effective environments from 2d PEPS. This is a key step in the simulation of 2d \emph{quantum} lattice systems. The same methods can also be considered for 2d \emph{classical} partition functions, and with some modifications they can also be used for 3d classical and quantum systems. { In fact, many of these methods were originally developed in the context of 2d classical statistical mechanics, see for instance the 1982 book by Baxter in Ref.~\cite{baxter}.} 

\bigskip 
\emph{(i) Boundary MPS:} this was the first method used to approximate the environment of a site on a 2d PEPS, both for finite \cite{PEPS} and infinite \cite{iPEPS} systems. The idea is to contract the full 2d lattice starting from a boundary (either an actual boundary for finite systems, or placing it manually at infinity for systems in the thermodynamic limit), with boundary tensors forming an MPS. The algorithm proceeds by contracting 1d MPOs made of lattice tensors into the MPS, mimicking a non-unitary time evolution that can be treated, e.g., with the techniques from Ref.~\cite{canonical}. For a PEPS of bond dimension $D$ and a boundary MPS of bond dimension $\chi$, the computational cost of this scheme for a 2d square lattice is $O(\chi^3 D^6 + \chi^2 D^8)$ for a horizontal/vertical boundary, and $O(\chi^3 D^4 + \chi^2 D^6)$ for a diagonal boundary \cite{iPEPS, tn}. 

\bigskip 
\emph{(ii) Corner Transfer Matrices (CTM):} this is a popular approach nowadays because it is relatively easy to implement and produces good-quality results. Focusing on a 2d square lattice of tensors, the idea is to find renormalized (coarse-grained) approximations to the tensors amounting to the contraction of all the tensors on the corners. Such tensors are the CTMs, which are well-known objects in the context of exactly-solvable statistical mechanical models. There are several schemes for dealing and finding such CTMs \cite{baxter, CTMRG, ourCTM, ghentCTM}. The computational cost depends on the specifics of the implementation. To have an idea, the method from Ref.~\cite{ourCTM} has a cost of $O(\chi^3 D^6)$ for a PEPS of bond dimension $D$ and a CTM of bond dimension $\chi$.  

\bigskip 
\emph{(iii) Tensor coarse-graining:} in this approach the main idea is to coarse-grain the network, by finding new renormalized tensors amounting for a ``zoom-out" and which take into account the main features of the TN at long distances. This is similar to a Kadanoff blocking for classical statistical mechanical models. Many schemes fall into this category, each one with its own advantages and drawbacks: Tensor Renormalization Group (TRG) \cite{trg}, Second Renormalization Group (SRG) \cite{srg}, Higher Order TRG (HOTRG), Higher Order SRG (HOSRG) \cite{hotrg}, Tensor Entanglement-Filtering Renormalization (TEFR) \cite{tefr}, Tensor Network Renormalization (TNR) \cite{tnr}, loop-TNR  \cite{loopUP} and TNR$_+$ \cite{tnrplus}. The specifics of the implementation for each one of these cases is different. However, the idea is that one typically contracts tensors, defining a new lattice in terms of some new tensors that amount for the contraction. The new tensors are then renormalized by truncating their bond indices with some isometries. The prescription for how to find such isometries is what defines the different schemes. Some schemes (such as TNR) also remove local entanglement before blocking the tensors.  The computational cost of each scheme can also be very different depending on the implementation. As an example, for a classical partition function with a TN structure of bond dimension $\chi$ on a square lattice, the SRG method has a cost of $O(\chi^{10})$ \cite{srgcost}. 

\bigskip 
\emph{(iv) Nested tensor network \cite{ntn}:} this approach is not a contraction scheme in itself (as the previous cases), but rather the idea of projecting the tensors of a 3d TN on a 2d plane, so that the resulting 2d TN can be contracted by any of the three strategies (i) -- (iii) above. An example is the TN for the norm of a 2d PEPS with bond dimension $D$, where bra and ket tensors are shifted with respect to each other, thus producing a new 2d TN similar to a 2d partition function with bond dimension $D$. Such a 2d TN can thus be contracted more efficiently than the one obtained with a double-layer approach, which has bond dimension $D^2$.

\subsection{Methods to obtain 2d states} 

Let us now focus on techniques to obtain TN states for 2d quantum lattice systems. As we shall see, in some cases the techniques explained before for 1d systems and 2d TN contractions will turn out to be fundamental. 

\bigskip 
\emph{(i) 2d DMRG:} people have considered using DMRG also to study 2d systems \cite{2dDMRG}. Even if, by construction, DMRG produces an MPS and is therefore a priory better-suited to deal with gapped 1d local Hamiltonians, the technique has also seen a lot of success in 2d because of its efficiency. The idea behind 2d DMRG is to have a stripe, or wrap the 2d system around a cylinder, and then use MPS as an ansatz for the 2d lattice following a snake pattern. The true 2d properties of the system are recovered by doing careful finite-size scaling with the thickness of the cylinder or the width of the stripe. This approach has been very successful in determining properties of the 2d $t-J$ model \cite{tjdmrg} as well as the spin liquid nature of the ground state of the Kagome Heisenberg Antiferromagnet (KHAF) \cite{khafdmrg}. An interesting evolution of 2d DMRG is to combine position and momentum basis for both directions, showing better performance in some situations \cite{mixedDMRG}. Still, the computational cost of 2d DMRG is eventually doomed for large 2d systems due to an exponential entanglement wall in the transverse direction that cannot be handled by an MPS with finite bond dimension.  

\bigskip 
\emph{(ii) 2d TTNs:} TTNs have also been used to study 2d systems, since their structure can be easily adapted for such geometries \cite{2dTTN}. Still, and as in the 2d DMRG case, the underlying TN structure has inherently 1d built-in correlations because of the absence of loops in the network, which again implies exponential entanglement walls and therefore only relatively good accuracy depending on the system and regime. Still, they can be pretty useful since their associated algorithms are quite efficient (as in 1d). For instance, they have been used to study confinement / deconfinement transitions of 2d $\mathbb{Z}_2$ lattice gauge theories \cite{2dZ2TTN}. 

\bigskip 
\emph{(iii) PEPS:} numerical algorithms based on PEPS are well-suited to tackle 2d systems. A PEPS has inherently-built truly 2d correlations and, as such, is a natural ansatz to study a wide variety of 2d systems. People have considered the ansatz for finite-size systems, the so-called finite-PEPS \cite{PEPS, finitePEPS}, but also for infinite systems, the so-called infinite-PEPS (iPEPS) \cite{iPEPS}. There are different ways of optimizing the tensors of a PEPS to obtain approximations of ground states. For instance, variational updates (both for finite \cite{finitePEPS} and infinite \cite{variPEPS} PEPS), as well as imaginary-time evolution via simple \cite{simpleUP}, full \cite{iPEPS} and fast-full \cite{ffUP} updates. Simple updates are very efficient but not necessarily accurate, while full and fast full updates take into account the effect of the environment when optimizing a PEPS tensor and are therefore slower but more accurate. Such environments are usually computed via the renormalization methods to contract 2d TNs explained in the previous section, i.e., using boundary MPS, CTMs, and tensor coarse-graining. Furthermore,  recently it was also shown how to compute excited states with PEPS for 2d systems \cite{exPEPS}, and how to do accurate extrapolations in the bond dimension \cite{extrapol}. The computational cost of PEPS algorithms heavily depends on the type of algorithm chosen. For instance, for a square lattice, simple-update algorithms with a mean-field environment \cite{gPEPS} have a cost of $O(D^5)$, whereas full and fast full updates have a cost of $O(\chi^3 D^6)$ (and with a prefactor that could be large), with $\chi$ and $D$ respectively the environment and PEPS bond dimensions. 

\bigskip 
\emph{(iv) 2d MERA:} The MERA has also been used as a variational ansatz to approximate ground states of 2d quantum lattice systems, see for instance Refs. \cite{MERA2dIs, MERAKag, MERA2dferm}. The computational cost of the approach strongly depends on the type of lattice as well as the specific choice of unitaries and disentanglers. For instance, the approach in Ref.~\cite{MERA2dIs} for an infinite square lattice has a computational cost of $O(\chi^{16})$, with $\chi$ the MERA bond dimension. 

\subsection{Combined methods} 

There have also been developments where TNs have been combined with other existing methods, or where TNs have been useful to understand other existing techniques. Some of these developments are briefly sketched in this section. 

\bigskip 
\emph{(i) Monte Carlo TNs:} Monte  Carlo methods have been used together with TN techniques in several ways. For instance, people have implemented Monte Carlo sampling to do variational  optimizations over TNs, as well as approximate calculations of effective environments \cite{MonteCarloTN1}. In the context of string-bond and plaquette-entangled states, Monte Carlo has been used also as a sampling technique for optimization and expectation value calculation  \cite{StringPlaquette}. The combination with Monte Carlo allows in principle to reach higher bond dimension in the calculations, at the cost of the length of the sampling. For instance, the cost of these methods when combined with MPS for periodic boundary conditions is typically $O(N \chi^3)$, with $\chi$ the MPS bond dimension and $N$ the length of the (finite) 1d system, and with a prefactor that depends on the number of samples.  

\bigskip 
\emph{(ii) TNs for Density Functional Theory (DFT):} DFT is one of the most popular numerical approaches to perform ab-initio calculations of real materials and molecules. In this context, TNs (and specially MPS) were used to produce systematic approximations to the exchange-correlation potential of electronic systems \cite{DFTTN}. 

\bigskip 
\emph{(iii) TNs for Dynamical Mean-Field Theory (DMFT):} TNs have also found important applications in DMFT. In particular, in a series of works \cite{TNDMFT} it has been shown that MPS techniques can be used as a high-accuracy and low-cost impurity solver, including applications to non-equilibrium systems. 

\bigskip 
\emph{(iv) TNs and wavelets:} recently, a number of connections have also been stablished between wavelet transformations and TNs. More specifically, in Ref.~\cite{WaveER} it was shown how Daubechies wavelets could be used to build an analytic approximation to the ground state of the 1d critical Ising model which, in turn, correspond to instances of the 1d MERA. Additionally, Ref.~\cite{WaveQC} showed how the structure of wavelet transformations adapts to that of a quantum circuit, and Ref.~\cite{WaveFerm} showed how the ground state of some fermionic systems could be understood via entanglement renormalization also using the language of wavelets. 

\bigskip 
\emph{(v) More synergies:} while the most relevant ``combined" methods are sketched above, there have been further developments in other directions which are also worth mentioning. For instance, the so-called Entanglement Continuous Unitary Transformations (eCUT) \cite{ecut} showed how to mix the idea of continuous unitary transformations and Wegner's flow \cite{cut} with TNs, by truncating the flow equation in its operator-entanglement content. Moreover, TNs have also been used in combination with perturbation theory, e.g., in Ref.~\cite{perpeps} it was shown how to construct an exact 2d PEPS up to a given order in perturbation theory. Another fruitful combination has been that of TNs and the randomized SVD \cite{rSVD}, which is proven to improve the efficiency of numerical algorithms such as TBED, DMRG and TRG \cite{rSVDalg}. TN states have also been used to develop generalized Lanczos methods \cite{genlan}. Finally, TN states were also useful to understand the mathematical structure of  exactly-solvable systems, e.g., the algebraic Bethe ansatz \cite{tnbethe}, the fermionic Fourier transform \cite{tnfft}, the XY spin chain \cite{mpsxy}, and Kitaev's honeycomb model \cite{tnkitaev}. 

\section{The role of symmetries} 

In this section we briefly overview the effect of symmetries in TNs. We sketch why having a symmetry implies a tensor factorization, and its important consequences. We also comment some basic aspects of fermionic tensor networks as a special case, as well as gauge symmetries, topological order, and the classification of quantum phases of matter. 

\subsection{Global symmetries} 

The implementation of global symmetries in TN algorithms has been considered (at least) since the early years of DMRG, and it is well-known that it can lead to important computational advantages. People have exploited this fact specially in DMRG \cite{dmrgsym}, the 1d MERA  \cite{merasym}Ê and 2d PEPS \cite{pepssym}, both for abelian symmetries such as $U(1)$ particle-conservation, but also for non-abelian symmetries such as $SU(2)$ rotation-invariance. For further details we refer the reader to, e.g., Refs.\cite{merasym, weich, ours}, which provide in-depth overviews. In addition to numerical advantages, an important theoretical consequence is that the so-called \emph{spin networks} appear naturally from TN states with symmetries. Spin networks are used, e.g., in loop quantum gravity to describe quantum states of space  at a certain point in time \cite{spinnetloop}. In this sense, TNs with symmetries turn out to offer not only efficient numerical methods for complex quantum systems, but also a very intriguing connection between quantum entanglement and gravity.  

\begin{figure}
\centerline{\includegraphics[width=\columnwidth]{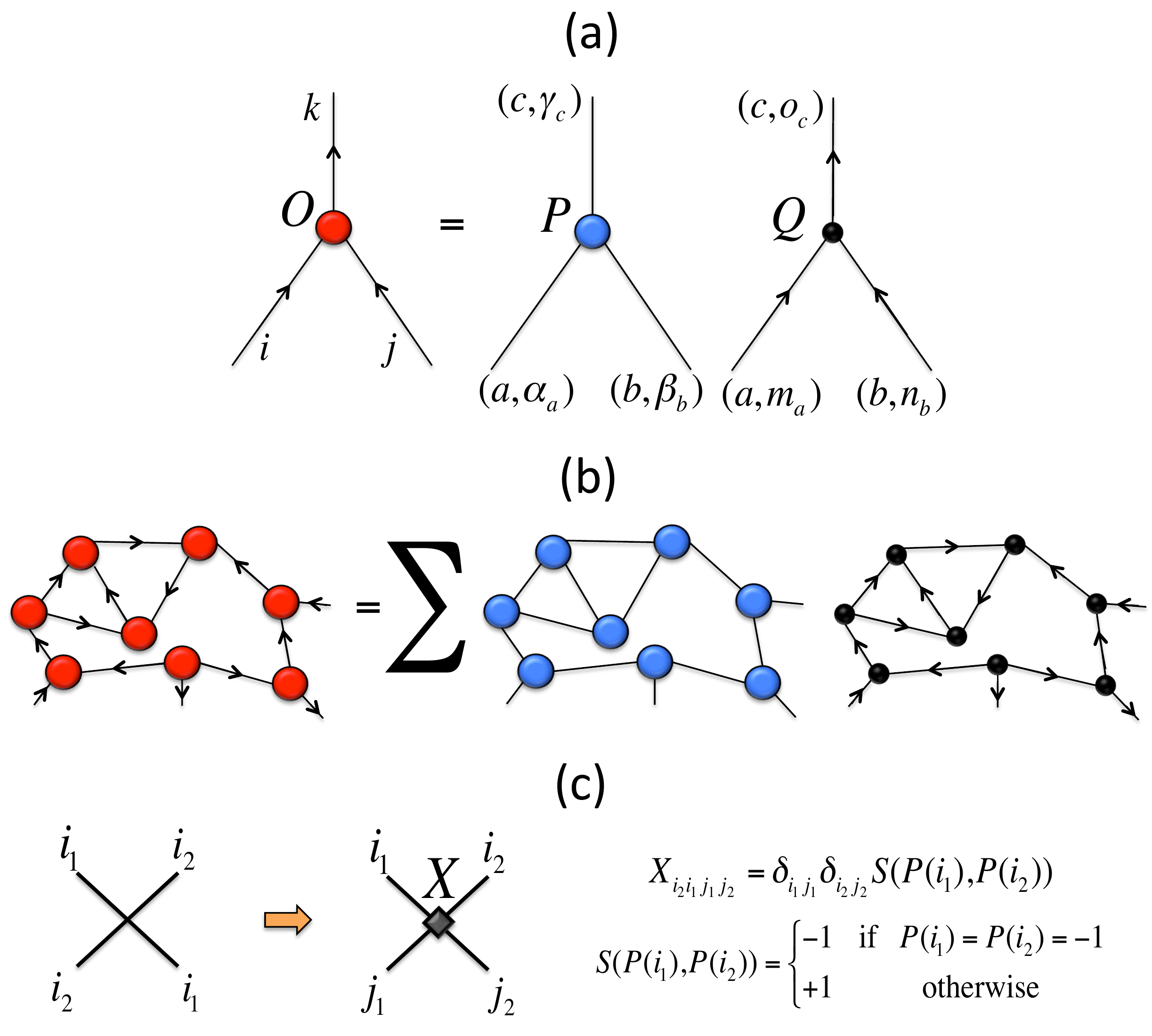}}
\caption{(a) A 3-index symmetric tensor $O$ decomposes into a degeneracy tensor $P$ and an intertwiner of the symmetry group $Q$ (which, e.g., for $SU(2)$ is a Clebsch-Gordan coefficient). (b) A symmetric TN decomposes as the sum of spin networks, with coefficients given by TNs of degeneracy tensors, and the sum being done over the degrees of freedom of internal indices. (c) In fermionic TNs, crossings between lines in diagrammatic notation are replaced by fermionic SWAP operators $X$. These operators take into account that the anticonmutation relation of fermionic operators in second quantization.}
\label{fig2}
\end{figure}

When dealing with local symmetries in TN states, Schur's lemma \cite{schur} ultimately implies that symmetric tensors can be decomposed in two pieces: one completely determined by the symmetry which acts on the subspaces of irreducible representations, and another that contains the actual degrees of freedom of the tensor, and which acts on a degeneracy subspace. As an example, for a symmetric tensor $O_{ijk}$ with three indices $(i,j,k)$, the Wigner-Eckart theorem implies that it can be decomposed as $O_{ijk} = (P^{abc})_{\alpha_a \beta_b \gamma_c} (Q^{abc})_{m_a n_b o_c}$, where   index $i$ decomposes as $i \equiv (a, \alpha_a, m_a)$ (and similarly for the rest),  $P^{abc}$ is a degeneracy tensor that contains all the degrees of freedom of $O_{ijk}$, and the structural tensor $Q^{abc}$ is completely fixed by the symmetry group. This is represented in the TN of Fig.~\ref{fig2}(a). In practice, this means that the symmetry constraint heavily reduces the degrees of freedom of the tensor. As a consequence, the whole TN also factorizes in two pieces: a TN of degeneracy tensors, and a TN of structural tensors. The second of the pieces is a spin network, see Fig.~\ref{fig2}(b). Thus, a symmetric TN for a quantum state $\ket{\Psi}$ of $N$ sites is a superposition of exponentially many spin networks with $N$ open indices, with coefficients determined by TNs of degeneracy tensors. 

There are many reasons why it is a good idea to implement symmetries in TN algorithms whenever this is possible. For instance, they allow to simulate systems with specific quantum numbers. But most importantly, symmetries also allow for a compact description and manipulation of the TN, thus leading to more efficient algorithms that can reach larger bond dimension.

\subsection{Fermionic systems} 

TN methods can also be used to simulate fermionic systems in any dimension, and directly in second quantization language. Fermionic statistics can be implemented in TN algorithms in different but equivalent ways \cite{MERA2dferm, fipeps, fPEPS, pit, foc, epjb}. The graphical implementation from Ref.~\cite{fipeps} is perhaps the simplest one to describe. This is based on two ``fermionization" rules: (i) use parity-symmetric tensors, and (ii) replace crossings in the planar representation of the TN by fermionic SWAP gates. The first rule is justified because fermionic parity, i.e., whether the total number of fermions is even or odd, is a good $\mathbb{Z}_2$ symmetry for fermionic systems. The second rule is also justified because fermionic SWAP gates, defined as in Fig.~\ref{fig2}(c), take correctly into account the anticommutation of fermionic operators in second quantization. In the end, fermoinic TN algorithms can be programmed in the same way as their bosonic counterparts, but taking into account these two simple rules, which only imply a subleading increase in the computational cost of the algorithm. 

\subsection{Gauge symmetries} 

Gauge symmetries can also be implemented naturally in the framework of TNs, leading to a similar TN decomposition as for global symmetries, but slightly different due to the local (gauge) nature of the symmetry \cite{gtn, gaugeMPS}. Many works have implemented the formalism of gauge-invariant TNs, focusing mostly on 1d systems and sometimes in 2d. In particular, $\mathbb{Z}_2$ lattice gauge theories (LGT) in (1+1)d have been considered with DMRG \cite{sugihara}. For the Schwinger model, i.e., quantum electrodynamics (QED) in (1+1) dimensions,  DMRG (without MPS formulation) was considered in several works \cite{dmrgOld}, whereas  MPS simulations have been done to compute the chiral condensate \cite{SchwingerDMRG} as well as thermal properties \cite{SchwingerThermal}, the mass spectrum \cite{SchwingerSpectrum}, the Schwinger effect \cite{SchwingerSchwinger}, the effect of truncation in the gauge variable \cite{SchwingerTruncation}, the case of several fermionic flavours \cite{SchwingerMultiflavour}, and the thermodynamic limit { of DMRG and its possible extrapolation to (2+1) dimensions} \cite{SchwingerKai}.  The consequences of gauge symmetry in the MPS of the Schwinger model was first elaborated in Ref.~\cite{gaugeMPS}. A gauge-invariant MPS ansatz was also used to compute the confining potential \cite{SchwingerPot} as well as the scattering of two quasiparticles \cite{SchwingerScat}. TN simulations have also been implemented recently for non-abelian lattice gauge theories in (1+1)d \cite{nonabelianTNS}. For higher-dimensional systems, gauge-invariant TN ansatzs have also been proposed analytically \cite{gtn, u1peps, su2peps}, which can be used as variational wave-functions to study lattice gauge theories in (2+1)d. 

\subsection{Topological order and classification of phases} 

TN states are the natural language for topologically-ordered systems, which can in turn be understood using gauge symmetries. There are several developments along this direction, both from the perspectives of analytics and numerics. From the analytics point of view, it has been proven that eigenstates of string-net models \cite{stringnet} admit an exact tensor network description \cite{StringnetPEPS, StringnetMERA}, where tensors have specific gauge symmetries. This means that all non-chiral topologically-ordered 2d phases on a lattice admit a TN description. Moreover, it has also been shown that { certain PEPS for fermionic systems as well as for spin systems} can handle chiral topological order, albeit the corresponding parent Hamiltonians { (which can be obtained directly from the PEPS tensors)} have either gapless bulk excitations or long-range interactions \cite{chiralPEPS}. { It is therefore a theoretical challenge to understand the correct TN framework to describe chiral topological states with gapped bulk and short-range parent Hamiltonians.} From the numerical perspective, TN algorithms have been used to compute phase diagrams of topological systems under perturbations, e.g., using TTNs \cite{2dZ2TTN} as well as 2d PEPS \cite{PEPSTO} and DMRG on cylinders \cite{DMRGTO}. In addition to these developments, TNs have also proven to be an extremely useful tool in computing symmetry-protected topological order in 1d \cite{1dSPT}, representing fractional quantum Hall states \cite{fqh}, simulating anyonic systems \cite{anyons}, and describing { theoretically} topological quantum computation \cite{FrankTO} as well as symmetry-enriched topological order \cite{seto}. A separate set of results, but related to the developments on topological order, has concerned the { theoretical} classification of quantum phases of matter \cite{class1}, for which TNs have also been useful. For instance, MPS and PEPS were used to classify phases of 1d and 2d quantum spin systems \cite{class2, class3, class31}. The classification of fermionic topological phases has also been studied using fermionic MPOs and fermionic PEPS \cite{class4, class5, class6}. 

\section{Holography} 

Several notions related to holography also play a key role in TN states. Here we sketch briefly how TNs provide a natural bulk-boundary correspondence via the so-called entanglement Hamiltonians. Moreover, we will also comment (very descriptively) on the connection to { the geometry of space-time} in quantum gravity and the idea that space-time may emerge from quantum many-body entanglement.

\subsection{Entanglement Hamiltonians} 

\begin{figure}
\centerline{\includegraphics[width=0.65\columnwidth]{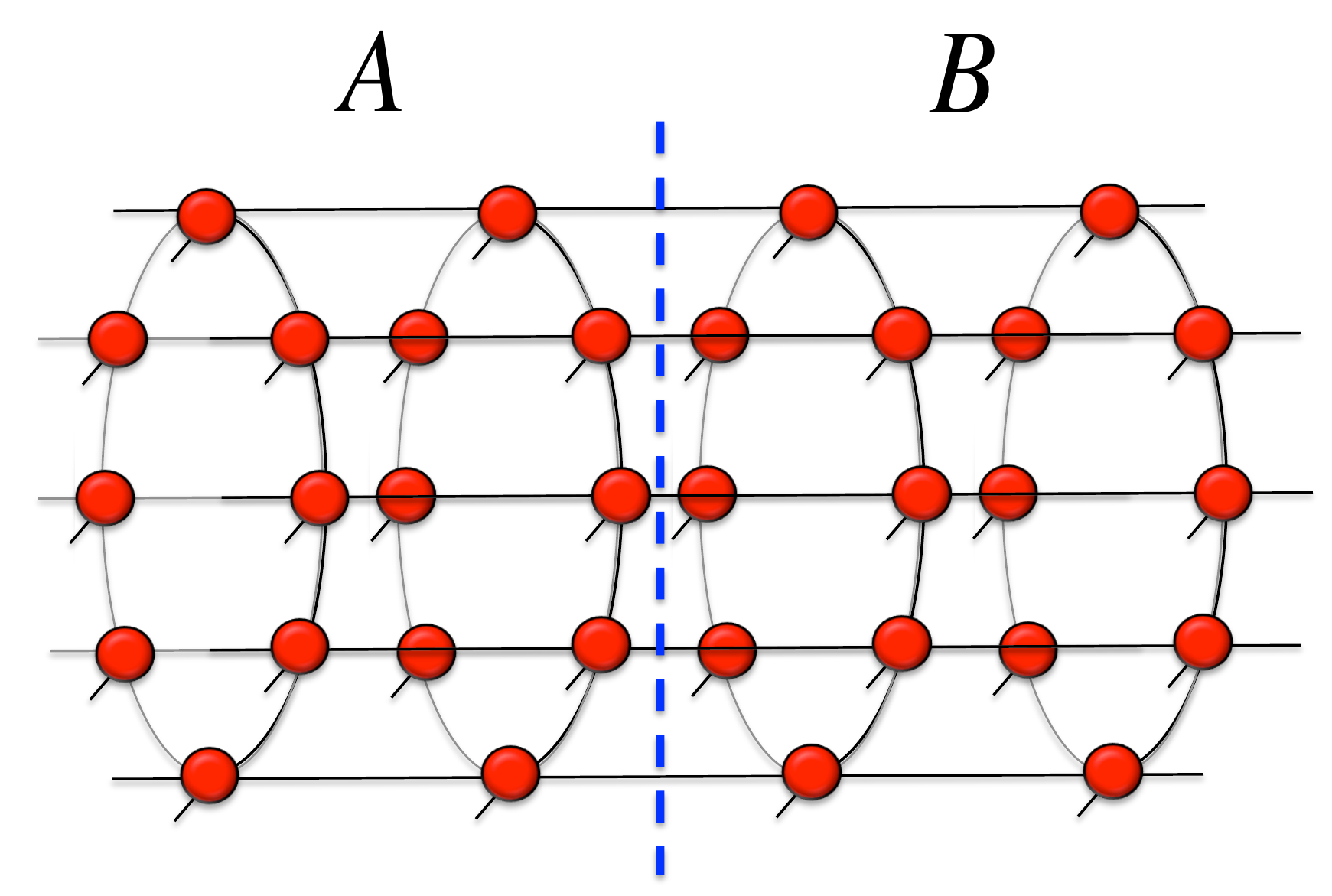}}
\caption{2d PEPS on an cylinder. To compute the entanglement Hamiltonian $H_E$, one makes a bipartition of the system ($A$ vs $B$), and computes the reduced density matrix $\rho$ of one of the subsystems using standard contraction techniques for PEPS (e.g., boundary-MPS with periodic boundary conditions). The eigenvalues of $\rho$ define then the eigenvalues of $H_E$, which can be grouped according to quantum numbers such as the vertical momentum (since there is translation invariance if we rotate the cylinder around its axis).}
\label{fig3}
\end{figure}

Li and Haldane \cite{liha} pointed out that the eigenvalues of the reduced density matrix of a bipartition codify important information about the boundaries of the system. Such reduced density matrix can be written as $\rho \propto e^{- H_E}$, with $H_E$ the entanglement Hamiltonian. The claim is that $H_E$ describes the fundamental degrees of freedom of the projection of the quantum state on a boundary. TNs turn out to be the natural arena to investigate such entanglement Hamiltonians for a variety of systems, and very specially for 2d PEPS \cite{enthampeps}. The main idea in such derivations is to wrap the 2d PEPS around a cylinder, and then study the entanglement spectra (i.e., the eigenvalues of $H_E$) of half of the cylinder versus the other half, see Fig.~\ref{fig3}. Such eigenvalues can be grouped in terms of their (vertical) momentum quantum number, and encode very useful information about the PEPS, such as possible gapless edge states and chiral  topological order \cite{chiralPEPS}. 

From the studies so far, there is an interesting correspondence. Consider PEPS that are ground states of 2d Hamiltonians with local interactions. It looks like, if the 2d system is gapped and not topologically ordered, $H_E$ is usually a 1d Hamiltonian with short-range interactions. However, for 2d critical systems, then $H_E$ is a 1d Hamiltonian with long-range interactions. Also, if the 2d system is gapped and topologically ordered, then $H_E$ is essentially a projector. All this is very interesting, because it justifies the numerical observation that  environment calculations in, e.g., infinite 2d PEPS converge quickly with very few iterations of a boundary MPS \cite{mpsgapped}. It seems, therefore, that 2d PEPS for ground states of  gapped 2d systems with no topological order can be contracted efficiently with good accuracy, even if this is not the case for a generic PEPS \cite{PEPSNPHard} or even on average \cite{PEPSaverage}. Some steps towards turning this observation into a mathematical theorem have already been taken \cite{DavidPEPS}.

\subsection{{ Emergent geometry}} 

\begin{figure}
\centerline{\includegraphics[width=\columnwidth]{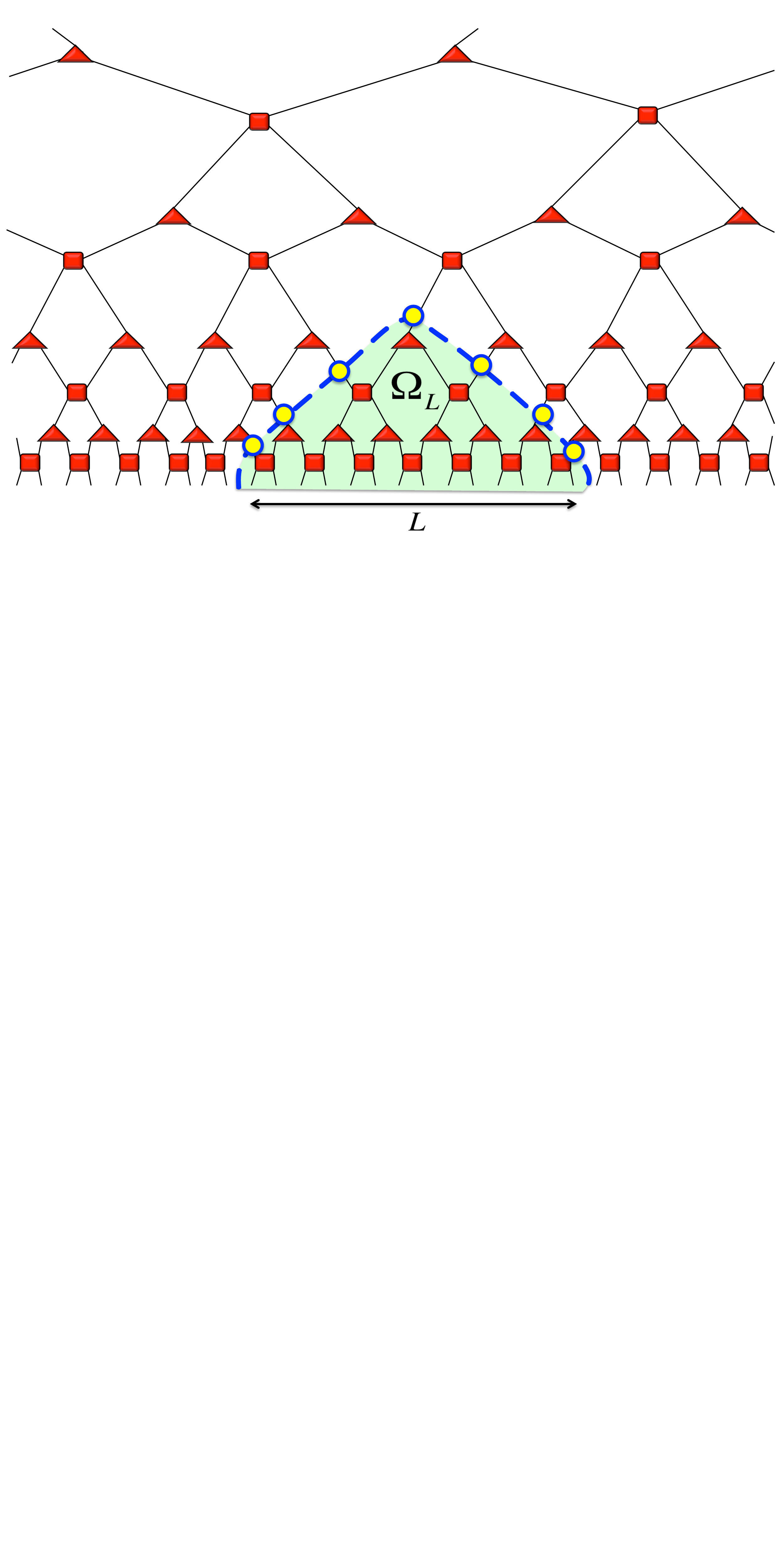}}
\caption{The entanglement entropy of a block of length $L$ for a 1d MERA is upper bounded as $S_L \le \log{\chi} \times \partial \Omega_L$, with $\partial \Omega_L$ the boundary of region $\Omega_L$ in the TN (i.e., the number of links crossed by the blue line) and $\chi$ the bond dimension. One quickly realizes that $\partial \Omega_L = O(\log L)$, and therefore one has that $S_L = O(\log L)$ for the 1d MERA. This calculation matches the behavior from CFTs in (1+1)d, and corresponds, precisely, to the lattice version of the Ryu-Takatanagi prescription to compute the entanglement entropy in AdS/CFT \cite{ryutakayanagi}. Entanglement is thus the \emph{area in holographic space} of the minimal surface separating the two regions. This is one of the key observations that motivates the analogy between MERA and AdS/CFT.}
\label{fig4}
\end{figure}

In the framework of holography, there is a very suggestive connection between entanglement, TNs, and quantum gravity: it looks like the MERA is a lattice realization of { a space with some geometry, where curvature is somehow linked to entanglement. The observation implies that space-time geometry may emerge from the underlying structure of entanglement in complex quantum states. An instance that has been studied in some detail is the possible relation between MERA and the  AdS/CFT or gauge / gravity duality \cite{adscft}.}  This connection between TNs and quantum gravity was originally noticed by Swingle \cite{swingle}, and later on investigated by several authors \cite{tngeom, Swingle2, Ryu, adsmera}. More specifically, for a scale-invariant MERA, the tensors in the bulk can be understood as a discretized AdS geometry, whereas the indices at the boundary correspond to the local Hilbert spaces obtained after a discretization of a CFT, see Fig.~\ref{fig4}. The connection can be made more formal by taking the continuum MERA \cite{cMERA} and evaluating the metric of the resulting smooth space in the bulk, with the curvature of the geometry being linked to the density of disentanglers \cite{Swingle2, Ryu}. As of today, the connection is very intriguing and has motivated a lot of research, specially from the string theory community. { In particular, there have also been claims that MERA does not actually correspond to and AdS geometry, but rather to a de-Sitter (dS) geometry \cite{dsgeo}. In a recent work, however, Milstead and Vidal showed that MERA is in fact neither AdS nor dS, but rather a \emph{lightcone} geometry \cite{lightcone}. In any case, and even if this connection is certainly suggestive and remarkable, the role played by TNs in the quantization of gravity is still unclear.} 

\section{Artificial intelligence}

In this section we will comment on the recent observation that neural networks (such as those used in deep learning) are in fact particular cases of TNs, as well as on the use of MPS to improve some methods of artificial intelligence. Additionally, we will also sketch the result that syntactic relations in language have a TN structure that is inherited by probabilistic language models.

\subsection{Machine learning}

Several promising connections between TNs and machine learning have been put forward recently. In Ref.~\cite{NNetTN} it was shown that deep learning architectures can actually be understood using the language of quantum entanglement. To name a couple of examples, convolutional networks correspond to specific cases of TTNs, and recurrent neural networks correspond to MPS. More generically, the whole machinery of quantum information and entanglement theory can be applied to understand neural networks in new ways. { One must however be careful, since in general neural networks are characterized by nonlinear functions, whereas TNs are linear and therefore obey the superposition principle.} In Ref.~\cite{bmachtn} it was also shown that there is an equivalence between restricted Boltzmann machines (a simple type of neural network) and TN states. { In Ref.~\cite{ignacio0} Boltzmann machines were also shown to be connected to some classes of TN states in arbitrary dimensions.} In addition to this, in Refs.\cite{miles1, miles2, maciej, han} it has been shown how MPS and TTNs can be used for supervised and unsupervised learning tasks of classifying images. Finally, in Ref.~\cite{miles3} it has been discussed how quantum circuits based on MPS and TTNs could be used to implement machine-learning tasks in near-term quantum devices, and in Ref.~\cite{ignacio} it has been explored how probabilistic graphical models motivate the concept of ``generalized TN", where information from a tensor can be copied and reused in other parts of the network, thus allowing for new types of variational wave-functions.    

\subsection{Language models}

From the perspective of computational linguistics, it has also been discovered recently that probabilistic language models used for speech and text recognition have actually a TN structure. This is a consequence of the fact that Chomsky's MERGE operation can be understood as a physical coarse-graining of information \cite{angel}. Such probabilistic models usually have the form of a TTN or even an MPS, i.e., loop-free TNs. In turn, this also matches the empirical observation that convolutional neural networks are quite good at language processing. In connection with the results commented in the previous section, it is clear that this is indeed so because such neural networks are TTNs, which encode the RG structure of language found in Ref.~\cite{angel}, and are therefore naturally well-suited for this task. 

\section{Further topics}

There are many other interesting results that have to do with TN states and methods, and which were not discussed so far. For obvious reasons we cannot summarize \emph{all} of them here. Nevertheless, here we sketch a few of them which we believe are worth pointing out.  

\subsection{2d Hubbard model}

The Hubbard model tries to capture the dynamics of electrons hopping on a lattice. In 2d, it is believed to be related to high-temperature superconductivity, but unfortunately its phase diagram has not been conclusively determined yet, even in the single-band approximation. In this context, it is worth mentioning that the best variational ground state energies computed so far for this model in the strongly correlated regime have been with TNs, specifically with the iPEPS algorithm for fermions \cite{extrapol}. Other related simpler fermionic models, such as the $t-J$ model, have also been simulated successfully using a variety of TN techniques, including 2d DMRG and iPEPS \cite{tjdmrg, tjTN}. 

\subsection{2d quantum antiferromagnetism} 

The antiferromagetic Heisenberg model on the Kagome lattice is the archetypical example of a frustrated magnet. Its ground state has remained elusive until recently, where 2d DMRG simulations unveiled that it is a quantum spin liquid \cite{khafdmrg}. Other TN methods have also attacked this problem, including 2d MERA with a specific disentangling structure \cite{MERAKag}, Projected Entangled Simplex States (PESS) \cite{PESSKag}, and iPEPS on coarse-grained lattices \cite{MarcPEPS}. Some of these simulations are compatible with a gapless quantum spin liquid ground state, but so far could not produce better energies than those obtained with 2d DMRG. Other models of quantum antiferromagnetism have also been studied via TNs. For example, iPEPS have been used extensively in the study of $SU(N)$ magnets \cite{frederic}.

\subsection{Conformal field theory} 

We discussed before the role of the MERA TN in the AdS/CFT correspondence. However, we would like to stress a number of recent results where TNs directly target properties of CFTs, not necessarily with the holographic duality in mind. In this respect, and always in the context of (1+1)d CFTs, Ref.~\cite{guif1} showed how to do the quotient of a MERA representation of the vacuum  taking it to a thermal state. Ref.~\cite{guif2} showed how to describe and coarse-grain the partition function of the 2d critical Ising model in the presence of topological conformal defects. Moreover, Ref.~\cite{guif3} studied how space-time symmetries are reflected in the cMERA of a free boson CFT, Ref.~\cite{guif4} explored the TN description of conformal transformations, and Ref.~\cite{guif5} proposed the interpretation of such TNs as path integrals on curved spacetime. In addition to this, and in connection to topological order, Ref.~\cite{cftto} investigated the mapping of topological quantum field theories to CFTs using 2d TNs. This family of results establishes a valuable dictionary between TNs and CFTs. 

\subsection{Quantum chemistry} 

TN numerical methods have also found important applications in quantum chemistry. For instance, DMRG has been used in this context already since some time ago \cite{chan}. The reordering of the fermionic orbitals has also been considered with MPS simulations \cite{ors1}. For a discussion on recent TN approaches to quantum chemistry, see also Ref.~\cite{ors2}.  

\subsection{Disorder and many-body localization} 

Several TN methods have been proposed to deal with disordered systems \cite{disordered}. Moreover, many-body localized (MBL) phases  have also been studied using tailored numerical TN methods. For instance, in Ref.~\cite{mbl1} a spectral tensor network was explored to represent all the spectrum of energy eigenstates. In Ref.~\cite{mbl2} a variational method over unitary MPOs was proposed to diagonalize MBL Hamiltonians. An alternative TN encoding of all eigenstates of MBL systems in 1d was proposed in Ref.~\cite{mbl3}. Finally, TNs were used to prove the robustness of MBL phases with SPT order \cite{mbl4}. 

\section{Outlook} 

This paper is an overview of developments around TN states and methods along different directions. We try to collect valuable information, including basic notions and bits of the overall current perspective. It is of course impossible to cover \emph{all} the developments, but hopefully we were able to summarize some of the key ones. { Importantly, notice that applications go well beyond quantum science. It is thus important to keep in mind that, in such cross-over applications (e.g., to artificial intelligence), typical properties of quantum mechanics are lost, such as unitarity. In such contexts one must therefore keep in mind that the TNs codify not a entanglement structure of a complex system, but rather the structure of its \emph{generalized} correlations, shall these be quantum or not.}

Our vision is that TNs will continue finding numerical and theoretical applications, both along established research directions, but also along new ones yet to be discovered. The lesson we learnt over the years is that  wherever one finds correlations, there may be a TN behind. And sometimes this leads to unexpected connections and pleasant surprises, pushing forward the boundaries of science. 

\bigskip 

{\bf Acknowledgements.-} I acknowledge financial support from Ikerbasque, DIPC and DFG, as well as discussions over the years with many people on several of the topics presented here. I also acknowledge M. Rizzi and P. Schmoll for a critical reading, and Iulia Georgescu for suggesting me to write this paper.  

\bigskip 
{ 
{\bf Glossary.-} 
\begin{itemize}
\item{Area-law: property by which the entanglement entropy of a region scales proportionally to the size of the \emph{boundary} of the region.}
\item{Parent Hamiltonian: Hamiltonian that has a given PEPS or MPS as unique ground state.}
\item{Gap: energy difference between the lowest-energy eigenstate of a Hamiltonian and the first excited state.}
\item{Correlation length: nonmathematically, this is the length scale at which correlations are sizeable in a many-body system.}
\item{Global symmetry: an operation that leaves invariant the system, and which acts equally on the whole system.}
\item{Gauge symmetry: an operation that leaves invariant the system, and which has the freedom to act differently at every point in the system.}
\item{Topological order: a type of order in quantum matter entirely due to global entanglement properties, and which does not exist classically. Other characterizations: excitations are anyonic, the topological entanglement entropy is non-zero, ground states are topologically degenerate, reduced density matrices are locally equivalent...}
\item{Chiral system: a system that breaks time-reversal symmetry.}
\item{Renormalization: the process of removing degrees of freedom that are not relevant to describe a complex system at different scales of some physical variable (energy, length...).}
\item{AdS space: a geometric space with negative curvature.}
\item{CFT: a quantum field theory with conformal symmetry, which includes scale invariance. Low-energy field theories of quantum critical systems are usually CFTs.}
\item{Neural network: a mathematical structure resembling the brain, made of a concatenation on non-linear functions (neurons) according to layers, and which can be optimized (trained) to recognize patterns, images, etc.}
\item{Boltzmann machine: a specific type of neural network where the target is to reproduce some Gibbs thermal probabilities.}
\item{MERGE: linguistic operation introduced by Noam Chomsky, which picks up two entities (e.g, noun and adjective) and produces a new one from the two (e.g., noun phrase).}
\item{Many-body localization: property of interacting quantum many-body systems with disorder leading to a phase of matter which does not self-thermalize.}
\end{itemize}
}

\end{document}